\documentclass{aa}  

\usepackage{graphicx}
\usepackage{txfonts}
\usepackage{bm}

\newcommand{\Eq}[1]{Eq.\ \ref{eq:#1}}
\newcommand{\Fig}[1]{Fig.\ \ref{fig:#1}}

\newcommand{\Figure}[1]{Figure\ \ref{fig:#1}}

\begin{document} 

\title{Abundance Effects from Protoplanetary Disk Outflows}
\author{{\AA}ke Nordlund}

   \institute{ Niels Bohr Institute, Jagtvej 155, 2100 Copenhagen, Denmark\\
    \email{aake@nbi.ku.dk}}
   \date{\today}

\abstract
% Context
{Systematic abundance differences that depend on the condensation temperatures of elements have been observed, in particular for stars similar to the Sun; `solar twins' and `solar analogues'. Similar differences have also recently been shown to exist between solar abundances and abundances of refractory elements in primitive chondrites. Numerous mechanisms have been proposed to account for these effects, including also differences observed in binary systems.   Some of the proposed explanations involve planet formation or planet destruction, potentially offering spectroscopic diagnostics of the presence of planets around stars.}
% Aims
{Rather than relying on specific mechanisms, this paper aims to show that the observed effects are a natural and unavoidable outcome of the star formation process itself, in which the associated outflows (winds and jets) carry away material, with efficiency varying with condensation temperature}
% Methods
{By using analysis based on modeling results and scaling laws, the trends and magnitudes of the effects are investigated, in three contexts: 1) with respect to differences between the Sun and solar twins, 2) with respect to differences between the Sun and CI-chondrites, and 3) with respect to differences between members of binaries.}
% Results
{It is shown that protoplanetary disk outflows indeed are expected to have differential abundance effects, with trends and magnitudes consistent with observations. The qualitative as well as semi-quantitative character of the effects are reproduced, in all three contexts.}
% Conclusions
{The results indicate that the observed systematic differences are likely due to the disk outflows associated with the accretion process, and are thus not directly related to planet formation.  In contrast to mechanisms relying on the tiny mass of planets leaving an observable signature, outflows carry away masses similar to the entire mass of the star, thus much more easily resulting in differential effects of the observed magnitudes, without having to assume that the abundance differences are limited to the convection zones of the stars.}

\maketitle

\section{Introduction}
Observations of young stellar objects, as well as high-resolution adaptive-mesh-refinement numerical simulations of star formation  have shown that outflows are always associated with star formation \citep{Pudritz2006,Fendt2007,2014IAUS..299..131N,2017ApJ...846....7K}.   As protoplanetary disks channel the accretion flows from the surroundings of newly formed star down to the star, mass can only reach the star by shedding angular momentum and potential energy; a process that relies mainly on transporting away the excess angular momentum and energy via the magnetic fields that are dragged in along with the gas and dust.  The magnetic fields connect the disks with the surrounding interstellar medium, and transport angular momentum and energy via Maxwell and Reynolds stresses, both directly, via electromagnetic Poynting flux, and indirectly, via the systematic outflows (disk winds and jets) driven by the magnetic field and assisted by the dissipation of magnetic energy into heat.  The outflows that allow accretion of part of the mass to occur return the rest of the gas and dust to the interstellar medium.   Volatile elements are more likely to become part of the ejected material while refractories are more likely to remain in the disk and become accreted to the star, and hence this process may be expected to leave an abundance imprint on the accreted material, with a general overabundance of refractory elements. The outflow process is therefore an excellent candidate to explain the systematic chemical abundance differences between the Sun and CI-chondrites \citep{2021A&A...653A.141A}, and among ‘solar twins’ \citep{2009ApJ...704L..66M,2010A&A...521A..33R,2015A&A...579A..52N,2017PhDT.......367B,2018A&ARv..26....6N,Katsova2022,Liu2022} and 'solar analogues' \citep{2018ApJ...865...68B,Youngblood2020,2024ApJ...965..176R,2025ApJ...980..179S}. 

As also shown by both observations and simulations, no two stars are subjected to exactly the same conditions when they form; important properties such as upstream specific angular momentum and average magnetic flux density vary from case to case, including between stars that form binaries.  The differential outflow mechanism thus also offers a natural mechanism to explain the observed differences in abundance patterns between the components of binary systems \citep{Desidera2004,Bonanno2006,2015ApJ...808...13R,Liu2021,2018ApJ...854..138O}.

Below, in Section \ref{sec:2}, the basic physical principles involved in the mechanism are introduced, while in Section \ref{sec:3} a simple, semi-analytical disk model is used to allow quantitative predictions of the effects.  In Section \ref{sec:4} the predictions are compared with the observed solar twin systematics, the observed Sun -- CI-chondrite differences, as well as with observed differences in binary systems.   The findings are discussed and summarized in Section \ref{sec:5}.

\section{Basic physical principles}\label{sec:2}
\newcommand{\SUB}[1]{_{\mbox{\small #1}}}
\def\Mdot{\dot{M}\SUB{acc}}
\def\MdotOut{\dot{M}\SUB{out}}
\def\MdotStar{\dot{M}_{*}}
\def\fStar{f_{*}}
\def\Mstar{M_{*}}
\def\Mint{M\SUB{inner}}
\def\Mouter{M\SUB{outer}}
\def\Mout{M\SUB{outflow}}
\def\Mtot{M\SUB{tot}}
\def\Mfrac{M\SUB{frac}}
\def\Fstar{F_{*}}
\def\Fint{F\SUB{inner}}
\def\Fouter{F\SUB{outer}}
\def\Fout{F\SUB{outflow}}
\def\Ffrac{F\SUB{frac}}
\def\SigmaMMSN{\Sigma\SUB{disk}}
\def\vAcc{v\SUB{acc}}
\def\vK{v\SUB{K}}
\def\cS{c\SUB{sound}}
\def\vOut{v\SUB{out}}
\def\Qacc{Q\SUB{acc}}
\def\Qrad{Q\SUB{rad}}
\def\sSt{\sigma\SUB{S}}
\def\hDisk{h\SUB{disk}}
\def\dDisk{\rho\SUB{disk}}
\def\dOut{\rho\SUB{out}}
\def\vOut{v\SUB{out}}
\def\fOut{f\SUB{out}}
\def\TDisk{T\SUB{disk}}
\def\alfven{Alfv\'en}
\def\rAcc{a\SUB{acc}}
\def\rVol{a\SUB{vol}}
\def\rRef{a\SUB{ref}}

Consider first of all the general mass balance of protoplanetary disks with outflows, noting that in this context it is not important whether the outflows are in the form of disk winds or jets.  What is important is only that outflows in both cases are closely linked to disk rotation, and typically have outflow velocities of order of the Kepler speed at the base of the magnetic field lines that support and drive the outflows.  This generic relation is understandable by considering the energies involved; it corresponds to the kinetic energy in the outflow being of comparable magnitude to the difference in the gravitational potential that the mass must overcome.
The loss of energy and angular momentum related to the outflow is also necessary to allow the remaining fraction of gas to be accreted.  

The energy loss can be further divided into the kinetic energy of the outflow, the electromagnetic (Poynting-) flux of energy associated with the outflow, and the associated dissipation of magnetic and kinetic energy into heat.
In the absence of more detailed modeling, one may assume that these energy dissipation channels are of similar magnitude, and in particular one may assume that a reasonable estimate of the disk surface temperature may be obtained by equating the radiative energy loss from the disk surface with the work done by gravity on the accreting gas.  Further details are given below, here only the basic result that temperature according to this model (and all other similar models) increases monotonically with decreasing radius is needed.

The fate of volatile, less volatile, and refractory elements is influenced and controlled by the increasing temperature, following the flow of accreting gas and solids. The drift of solids with respect to the gas can be neglected---this is justified by noting that the bulk accretion speed at 1 AU for the Minimum Mass Solar Nebula \citep[MMSN --][]{1981PThPS..70...35H} at an accretion rate of $10^{-5}$ solar masses per year is of order 250 m/s and thus greatly exceeds the radial drift speed, which is at most of order the head-wind (typically less than 100 m/s).

Species with different condensation temperatures will sublimate at temperatures near their condensation temperatures, and once in the gas phase they will participate in the outflows. 
The outflow, at any one radius, thus consists of the dominant volatile elements---primarily hydrogen and helium---plus the species that have condensation temperatures below the disk surface temperature at the given radius. 
More refractory species, with condensation temperatures above the local disk surface temperature, will remain in solid form, and since a fraction of the more volatile elements (including the dominant ones) have been removed by outflow further out in the disk, the relative abundances of the more refractory elements increases with decreasing radius.

Let $\rAcc$ be the radius inside which all elements are in gas form, with $\Mint$ the mass loss occurring there.  That mass loss does not by itself change the fractional elemental abundance of the star, since the relative abundances in that mass equals that of the accreting material.  The mass loss is nevertheless important, since it contributes to (and most likely dominates) the total mass loss of the system.

Correspondingly, denote by $\rVol$ a low temperature radius limit, outside of which practically no sublimation occurs, and outside of which the outflow consists of mainly hydrogen and helium.  The radius interval outside of $\rVol$ thus also does not contribute to change the relative element abundances, except by reducing the amount of hydrogen and helium by an amount denoted $\Mouter$, and thus uniformly increasing the other element abundances.  

Applying mass conservation, and denoting the total accretion of mass to the disk $\Mtot$, one can write
\begin{equation}
    \Mtot = \Mstar + \Mout = \Mstar + \Mint + \Mfrac + \Mouter \,,
\end{equation}
where $\Mfrac$; the mass lost in the radius interval $\rRef < r < \rVol$ is the crucial part of the mass balance, consisting of the integrated fraction of the outflow that contains varying amounts of refractory elements.  Normalizing on $\Mtot$, one can rewrite this as
\begin{equation}
    \Fstar + \Fout = \Fstar + \Fint + \Ffrac + \Fouter = 1 \,,
\end{equation}
It is generally assumed, as supported by evidence from both observations and numerical modeling, that the amount of mass lost in outflows is similar to the amount that reaches the star.  
This can indeed be supported by a simple lever arm argument:  The loss of angular momentum occurs when out-flowing gas is forced to co-rotate with angular velocity of the anchor points of the field lines they slide along; they thus acquire an excess angular momentum of the order the radius extent of that lever arm, relative to the radius of the anchor point, which is likely to be of order unity. Mass and angular momentum balance then results in a mass loss rate of similar magnitude as the net accretion rate.

The exact fractions are not important in the current context, and in any case the fractions certainly vary from case to case.  What is important for the arguments in the current context is to estimate how large or small the $\Ffrac$ fraction of the outflow might be, relative to the stellar fraction $\Fstar$.  To do so, one needs be more concrete with respect to the physical conditions in disks, and about the fundamental physical principles that determine them.

%%%%%%%%%%%%%%%%%%%%%%%%%%%%%%%%%%%%%%%%%%%%%%%%%%%%%%%%%%%%%%%%%%%%
\section{Semi-analytical scaling}\label{sec:3}
\begin{figure}[t]
\centering
\includegraphics[width=0.9\linewidth]{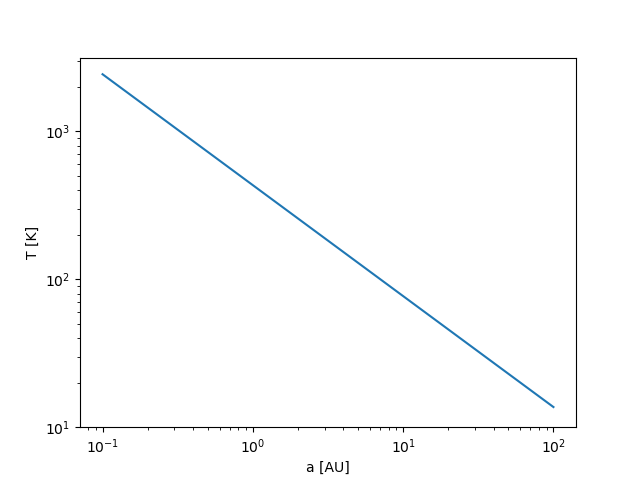}
\caption{\label{fig:T}Disk surface temperature according to \Eq{T}, for a solar mass stellar embryo with an accretion rate of $10^{-5}$ solar masses per year.}
\end{figure}

%%%%%%%%%%%%%%%%%%%%%%%%%%%%%%%%%%%%%%%%%%%%%%%%%%%%%%%%%%%%%%%%%%%%
As an instrument to reveal approximate scaling relations, with corresponding predictions of the qualitative and quantitative effects, one may use a standard disk model, similar to the MMSN model \citep{1981PThPS..70...35H}, where
%.....................................................................
\begin{equation}
    \Sigma = \Sigma_1\left(\frac{a}{AU}\right)^{p} \sim a^{p}\,,
\end{equation}
%.....................................................................
with $\Sigma_1 = 1700$ g cm$^{-2}$ and $p=-3/2$ in the standard MMSN case \citep{1981PThPS..70...35H}.
Assuming that the disk is optically thick, and that the accretion heating
%.....................................................................
\begin{equation}
    \Qacc = \frac{\Mdot}{2\pi a \Sigma} \frac{G_0 M_*}{a^2}
\end{equation}
%.....................................................................
is balanced by a disk surface cooling
\begin{equation}
    \Qrad = -\frac{2 \sSt T^4}{\Sigma} \,,
\end{equation}
one gets the estimate
%.....................................................................
\begin{equation}\label{eq:T}
    \TDisk = \left(\frac{G_0 \Mdot M_*}{4\pi\sSt a^3}\right)^{1/4} \,,
\end{equation}
%.....................................................................
illustrated in \Fig{T}, which shows the result for a case with an accretion rate of $10^{-5}$ solar masses per year, typical of the  time interval when solar mass stars accrete the first 50\% of their mass \citep{2017ApJ...846....7K}. 
Note that at times when the mass of the embryo was smaller, the relation was just shifted inwards in radius.  This has no effect on the arguments below.

%---------------------------------------------------------------------
%---------------------------------------------------------------------

If the disk surface density scales as $\Sigma \sim a^p$ then,
to conserve mass, the radial accretion velocity $v_a$ must scale as $v_a \sim a^{-(p+1)}$
and thus the time a Lagrangian mass element spends in a radius interval $d a$ scales as
\begin{equation}
dt \sim a^{(p+1)} da  
\end{equation}

Now, let $X_s = m_s/m_H$ stand for the abundance relative to hydrogen of some species of interest.
As discussed in Section 2, the change of the relative abundance of different species arises because the more volatile elements evaporate at larger orbital radii, and hence participate to a larger extent in the outflow, and therefore the relative abundance of more refractory elements increases, in proportion to the loss of Hydrogen. Following the Lagrangian inwards motion, the relative rate of abundance change thus scales as 
\begin{equation}\label{eq:dotX}
    \frac{\dot{X}_s}{X_s}
    = -\frac{\dot{m}_H}{m_H}
    \sim -\frac{\rho_o \, v_o}{\Sigma} \,
\end{equation}
where $\rho_o$ is the volume density at the point where the outflow speed $v_o=v_K$, the Kepler speed.  Then express the density $\rho_o$ as a fraction $f \sim a^q$ of the average disk volume density $\rho_d=\Sigma/2 H$. With the sound speed $c$ scaling with $T^\frac{1}{2}$, and where $H = a \, c/v_K$ is the disk scale height, the resulting scaling is
\begin{equation}\label{eq:dlnX}
    - d\ln X_s = 
    \sim \frac{f \, v_K}{H} dt
    \sim \frac{f \, v_K^2}{a \, c} dt
    %\sim a^q a^{-2} a^\frac{3}{8} dt
    \sim a^{(q - 2 + \frac{3}{8})} dt
    \sim a^{(p+q - 1 + \frac{3}{8})} da
    \,
\end{equation}
Integrating over $a$, from infinity (with zero abundance change) to the orbital radius $a_s$ where the species evaporates, one obtains the scaling of the logarithmic abundance differences,
\begin{equation}\label{eq:diff}
  \Delta\ln {X}_s
  \sim a_s^{p+q+\frac{3}{8}} = a_s^{-r}
  \sim T_s^{4 r/3}  \,,
\end{equation}
where the relation between temperature and orbital radius has been used to express the scaling in terms of the temperature $T_s$ where the species evaporates.
An estimate of a likely range $1 < r < 3$ may be obtained by adopting the MMSN scaling $p=-\frac{3}{2}$ and assuming $-2 < q < 0$, since it seems likely that the mass loading of the outflow increases with decreasing orbital radius.

To make predictions more quantitative would require detailed and costly 3-D non-ideal MHD modeling of protoplanetary disks with a level of detail currently not affordable, but the scaling predictions already make it possible to compare with observational trends.

%%%%%%%%%%%%%%%%%%%%%%%%%%%%%%%%%%%%%%%%%%%%%%%%%%%%
\section{Comparisons with observations}\label{sec:4}
The predictions put forward above can be tested against observations in three different contexts:
\begin{enumerate}
    \item Abundance differences between the Sun and solar twins that correlate with condensation temperatures are well documented since \cite{2009ApJ...704L..66M} and \cite{2010A&A...521A..33R}, with further results and references given in \cite{2018ApJ...865...68B} and \cite{2018A&ARv..26....6N}, and most recently by \cite{2024ApJ...965..176R} and \cite{2025ApJ...980..179S}.
    \item The predictions of solar abundance determinations on the one hand and abundance determinations for meteorites on the other hand are now sufficiently precise to explore systematic trends and differences; cf.\ \cite{2021A&A...653A.141A}.
    \item Binary components with similar temperatures offer another possibility to compare abundances differentially; cf. \cite{Desidera2004}, \cite{2015ApJ...808...13R}, \cite{2018ApJ...854..138O} and references therein.
\end{enumerate}
%---------------------------------------------------------------------
\begin{figure}[t]
\centering
\includegraphics[width=0.9\linewidth]{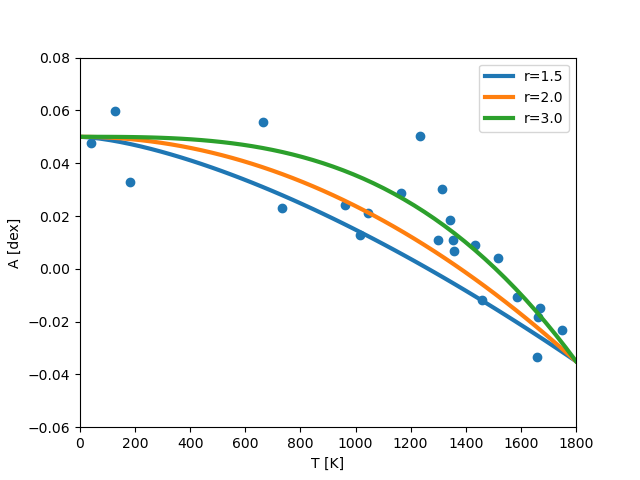}
\caption{\label{fig:solar-twins}Differential abundances of the Sun relative to an average of solar twins (dots), from Fig.\ 6 in \cite{2009ApJ...704L..66M}.  The curves shows predictions based on \Eq{diff}, with $r=1.5$ (blue), $r=2$ (orange), and $r=3$ (green).}
\end{figure}
%---------------------------------------------------------------------

Note that, for consistency with the original figures, the comparisons are made in terms of $A=\Delta\log_{10} X$ (dex) values, for which 0.1 dex corresponds to about 23\%, so even though the changes are small in terms of achievable accuracy in abundance measurements, in terms of fractions of stellar masses they nevertheless represent quite significant mass losses.

For simplicity, and since the observational error bars are significant (cf.\ the original figures), the fits to the observational trend have been made by eye, by adjusting the zero points and total amplitudes, and with the power index $r$ controlling the shape of the curves.

%%%%%%%%%%%%%%%%%%%%%%%%%%%%%%%%%%%%%%%%%%%%%%%%%%%%%%%%%%%%%%%%%%%%%%
\subsection{Abundance differences between the Sun and and solar twins}\label{sec:4.1}
When interpreting observed trends in comparisons between the Sun on the one hand and solar twins and solar analogues on the other hand it is important to keep in mind that this is a comparison between one member of an ensemble and the rest of the ensemble.  As shown above, the effect of outflows in general is to create an overabundance of refractory elements relative to volatile elements, but since circumstances under which solar mass stars form can vary greatly \citep{2017ApJ...846....7K}, the sample has a significant spread \citep{2018ApJ...865...68B}, and a comparison of a single sample with the average is thus likely to show a trend, which can go in any direction.

The first observations of such a trend among solar twins relative to the Sun were reported by \cite{2009ApJ...704L..66M}; their results are shown in \Fig{solar-twins}, together with normalized predictions based on \Eq{diff}.  As illustrated by Figs. 3 and 4 of \cite{2010A&A...521A..33R} their results, which include both solar twins and solar analogues, are consistent with those of \cite{2009ApJ...704L..66M}. 

As per the discussion above, the apparent under-abundance of refractories in the Sun can be interpreted as the average solar twin having a larger overabundance of refractories; in other words showing that the Sun was subjected to a smaller fractionation effect than the sample average.

The sharply increasing trend for large temperatures indicates a power index in the range -2 to -3 (orange and green curves), a conclusion that is further strengthened by Fig.\ 4 below.  It indicates that the outflow is dominated by temperatures larger than about 1000 K, a conclusion that is also consistent with the fact that the differential effect is only of order 0.1 dex, while other evidence suggests that the total mass loss in outflows can be as large or larger than the mass of the star.

%%%%%%%%%%%%%%%%%%%%%%%%%%%%%%%%%%%%%%%%%%%%%%%%%%%%%%%%%%%%%%%%%%%%
\subsection{Abundance differences between the Sun and CI chondrites}\label{sec:4.2}
The Sun is the only star for which direct measurement of the abundance differences between the primordial material, represented by CI-chondrites, and the final stellar abundances exists.   According to Fig.\ 6 of \cite{2021A&A...653A.141A}, the span of abundance differences is about 0.12 dex over the range from 400 K to 1800 K; similar in magnitude but opposite in sign relative to the difference between the Sun and solar twins (and with necessarily larger error bars since except for being anchored on an assumed common silicon abundance these are absolute measurements). 

The initial discovery that the Sun was under-abundant in refractories relative to most solar twins led to the idea that abundance measurements could be used to pin-point the presence or not of planets.  But the results by \cite{2021A&A...653A.141A} show that the Sun is actually slightly overabundant in refractory elements, relative to primitive CI chondrites, and the ``planet search idea''---already in conflict with the initially fully convective structure of early solar type stars---is therefore no longer viable.

\begin{figure}[t]
\centering
\includegraphics[width=0.9\linewidth]{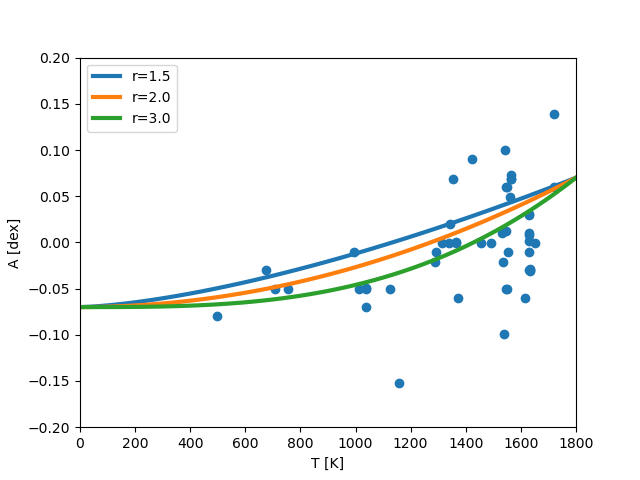}
\caption{\label{fig:fig3}Differential abundance profile for solar abundances minus CI-abundances, from Fig.\ 6 in \cite{2021A&A...653A.141A}, compared with predicted values from \Eq{diff}.}
\end{figure}

In relation to the idea pursued here, the comparison in \Fig{fig3} shows that the prediction first of all has the correct sign; a larger abundance of refractories in the Sun, relative to CI chondrites is reproduced.  As with the solar twin comparison, the largest differences occur for the highest temperatures.  The amplitude of the effect being similar to the difference between the Sun and solar twins indicates that as a sample from an ensemble, the Sun is a typical case with less-than-average outflow effects.
 
%---------------------------------------------------------------------
%---------------------------------------------------------------------

%%%%%%%%%%%%%%%%%%%%%%%%%%%%%%%%%%%%%%%%%%%%%%%%%%%%%%%%%%%%%%%%%%%%%%%
\subsection{Abundance differences between components in binary systems}\label{sec:4.3}
Components of binary systems that show differences in abundances that depend on condensation temperatures have been found \citep{2014ApJ...790L..25T,2015ApJ...808...13R,2017A&A...604L...4S,2017A&A...608A.112N,2018ApJ...854..138O}.  This has been regarded as a conundrum, and has been notoriously hard to explain by other proposed mechanisms.  With the mechanism proposed here, it follows directly from the diversity of conditions under which individual accretion disks form; there is no reason to expect components of a binary to have more similar accretion disk conditions than any other two stars.  The number of binary systems for which sufficiently accurate abundance measurements are available is less than in the case of solar twins, but the variability and range of condensation temperature slopes seem consistent \citep[cf.\ Fig.\ 8 in][]{2018ApJ...865...68B}.

\begin{figure}[t]
\centering
\includegraphics[width=0.9\linewidth]{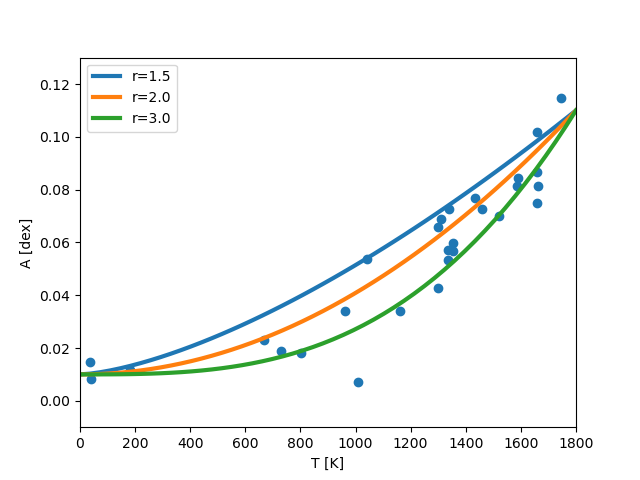}
\caption{\label{fig:Ramirez+2015}Abundance differences as a function of condensation temperature for two components of a binary system, from Fig.\ 5 in \cite{2015ApJ...808...13R}, compared with predicted values from \Eq{diff}.}
\end{figure}

\Figure{Ramirez+2015} shows a comparison of data from \cite{2015ApJ...808...13R} with the \Eq{diff} scaling, demonstrating that the scaling is consistent with observations also in this case, and that the magnitude of the effect is similar to the range of differences between solar twins, and the amplitude of the difference between the Sun and CI chondrites in \Fig{fig3}.
%---------------------------------------------------------------------
%---------------------------------------------------------------------

%%%%%%%%%%%%%%%%%%%%%%%%%%%%%%%%%%%%%%%%%%%%%%%%%
\section{Discussions and conclusion}\label{sec:5}
The three types of comparisons with observations are all consistent with the considerations in Section \ref{sec:2} and the predictions in Section \ref{sec:3}.   The general interpretation is that stars are subjected to differential removal of volatile and refractory material from their accretion disks, which causes a general overabundance of refractories in the stars, relative to the abundances in the protostellar cores from which they were formed.  This is directly supported by the observed differences between the solar abundances and the abundances in CI chondrites; cf.\ Section \ref{sec:4.2} and \cite{2021A&A...653A.141A}.  The positive curvature of the dependence on condensation temperature indicates that the integrated differential mass loss increases faster than in inverse proportion to orbital radius (cf.\ \Eq{diff}), corresponding to a local mass loss that increases faster then the inverse square of the orbital radius (cf.\ \Eq{dlnX}).

Since different stars---including stars that happen to end up with the same final mass---in general will have accretion histories and accretion disks that differ in properties such as accretion rate as a function of time, radial mass density profiles, as well as strengths and orientations of the magnetic fields that allow them to accrete, one must indeed expect differences in abundance profiles between individual stars with otherwise nearly identical final properties.   In particular, one must expect differences between the abundance profiles between the Sun and solar twins of the type found by e.g.\ \cite{2009ApJ...704L..66M} and \cite{2010A&A...521A..33R}.

Similar differences must also be expected between components of binary systems, since even if they accrete from essentially the same protostellar envelope, their accretion disks can have quite different properties.  This offers a simple and natural explanation to the differences found by for example \cite{2015ApJ...808...13R,2017A&A...608A.112N,2018ApJ...854..138O}, and others.

In general, more direct comparisons could only be made by modeling the exact circumstances of each case, which is in practice impossible, since one cannot access the exact accretion history in any of the cases.  However, the general magnitude of the differences are consistent with the assumptions put forwards in Section \ref{sec:2} and the formalism developed in Section \ref{sec:3}, since with total mass losses of order 50\% or more of the accreting material, it is to be expected that differential effects may reach several tens of percent in some cases, and that differences between otherwise nearly identical stars therefore also can reach similar values.

It should be emphasized that the differences predicted by the conceptual model proposed here applies to whole stars, while proposed explanations that rely on material deposited in or by planets generally are only viable if mass is taken to be relative only to the mass in the convection zones.  In particular, the difference between the Sun and CI chondrites cannot be explained at all in these models, since mass removed into planets early would cause an under-abundance in the Sun relative to CI-chondrites, and to at all be significant (but with still the wrong sign) planet formation would have to be assumed to be very much delayed, in conflict with current evidence of very early planet formation.

In conclusion, the observed abundance differences as a function of condensation temperature finds a natural explanation as a result of differential retention and removal of material in protoplanetary disk accretion and outflows. Even though it is merely a semi-analytical analysis, it reproduces several distinct observed properties, which hitherto have gone either unexplained, or have required specifically tuned explanations:
\begin{enumerate}
    \item The excess of refractories in the Sun, relative to CI-chondrites, which is otherwise hard to reconcile with the common notion of the Sun being ``refractory depleted'' \citep{2024ApJ...965..176R}.
    \item The difference of the Sun relative to solar twins, as a likely ``draw'' from an ensemble with randomly varying accretion properties.
    \item The abundance difference between members of binaries, including the frequency of occurrence.
    \item The order of magnitude of the abundance differences, which is consistent with a total mass loss that is a few times larger than the abundance difference amplitude.
    \item The similarity of the magnitude of the effect in the three different contexts, which would need to be seen as coincidences in scenarios with separate explanations, but comes out as an expected result in the current scenario.
\end{enumerate}

Specifically, with respect to the Sun and solar twins, the often expressed concern that the Sun is an `atypical solar mass star' becomes moot in the current context, since there is nothing strange with a single sample in a given distribution lying on one side of the ensemble average; on the contrary it would have been a rare (and unfortunate) coincidence if the Sun had happened to agree with the solar twin average.

\textit{Acknowledgments}:
I'm grateful to Bengt Gustafsson for numerous discussions on this subject, and for his initiative to drive this issue as a crucial one, at the intersection between stellar abundance determinations and the quest to understand star and planet formation.  Many thanks also to Paolo Padoan and Michael K{\"u}ffmeier, for proof reading and helping to improve the presentation,

\bibliographystyle{aa}
\bibliography{main}
\end{document}